\documentclass[unnumsec,namedate,webpdf,traditional,mediumone]{oup-authoring-template}

\usepackage{booktabs}
\usepackage{array}
\usepackage{tabularx}
\usepackage{longtable}
\usepackage{pdflscape}
\usepackage{ragged2e}
\usepackage{todonotes}
\graphicspath{{figures/}}
\usepackage{graphicx}
\usepackage{multirow}
\usepackage{placeins}

\newcolumntype{Y}{>{\RaggedRight\arraybackslash}X}
\newcolumntype{L}[1]{>{\RaggedRight\arraybackslash}p{#1}}

\input{supplementary_methods-labels.tex}
\input{study_evidence-labels.tex}

\begin{document}

\journaltitle{Biostatistics}
\DOI{To be assigned}
\copyrightyear{2026}
\pubyear{2026}
\access{Advance Access Publication Date: To be assigned}
\appnotes{Critical Review}

\firstpage{1}

\title[Inference under response-adaptive experimental designs]
{Statistical inference under response-adaptive experimental designs: a systematic review}

\author[1,$\ast$,$\dagger$]{Haochen Song}
\author[2,$\dagger$]{Joseph Williams}
\author[3]{Kiarash Kianidehkordi}
\author[4]{Audrey Durand}
\author[5]{Anna Rafferty}
\author[1]{Tong Li}
\author[1,$\ast$]{Meredith Franklin}

\address[1]{\orgdiv{Department of Statistical Sciences},
\orgname{University of Toronto},
\orgaddress{Toronto, \country{Canada}}}
\address[2]{\orgdiv{Department of Computer Science},
\orgname{University of Toronto},
\orgaddress{Toronto, \country{Canada}}}
\address[3]{\orgname{University of Toronto},
\orgaddress{Toronto, \country{Canada}}}
\address[4]{\orgdiv{D\'{e}partement d'informatique et de g\'{e}nie logiciel},
\orgname{Universit\'{e} Laval},
\orgaddress{Qu\'{e}bec, \country{Canada}}}
\address[5]{\orgdiv{Department of Computer Science},
\orgname{Carleton College},
\orgaddress{Northfield, \state{Minnesota}, \country{USA}}}

\corresp[$\ast$]{Corresponding authors: Haochen Song
(\href{mailto:fred.song@mail.utoronto.ca}{fred.song@mail.utoronto.ca})
and Meredith Franklin
(\href{mailto:meredith.franklin@utoronto.ca}{meredith.franklin@utoronto.ca}).\par}
\corresp[$\dagger$]{Haochen Song and Joseph Williams are co-first authors.}

\corresp[]{\par\medskip
\begin{minipage}{\textwidth}
\fontsize{9}{11}\selectfont\raggedright
\textbf{Funding and acknowledgment.} This work was supported by the National Science Foundation [grant number 2209819]. Meredith Franklin's contributions to this work were supported by the Natural Sciences and Engineering Research Council of Canada (NSERC) [grant number RGPIN-2024-06444]. The authors determined the review methods and framework, interpreted the findings, prepared the manuscript, and decided to submit it for publication independently of the funding organizations.
\par\smallskip
\textbf{Conflict of interest.} There is no declared conflict of interest.
\end{minipage}\par}

\received{Date}{0}{Year}
\revised{Date}{0}{Year}
\accepted{Date}{0}{Year}

\abstract{Response-adaptive experimental designs, including bandit-based approaches, modify treatment allocation as data accrue, often with the aim of improving participant benefit, allocation efficiency, or information about treatment outcomes and effects. Since treatment allocation may depend on accumulating outcomes and other trial information, the resulting data collection process can differ from that under conventional randomization, and analyses that ignore this dependence may produce biased estimates or fail to maintain nominal type I error or interval coverage. We conducted a systematic review of statistical inference under response-adaptive experimental designs in health and biostatistical research, examining the allocation designs studied, inferential procedures used, approaches to establishing inferential validity, and reported operating characteristics. Following PRISMA 2020 reporting guidance, we searched six bibliographic databases for peer-reviewed studies published between January 2010 and October 2025. Of 774 records identified, 36 studies met the inclusion criteria. We summarized the findings using study-level counts and structured narrative comparisons, examining the evidence supporting inferential findings in relation to the stated assumptions and evaluation settings. Twenty-five studies focused on developing or evaluating methods or designs, and six combined methodological work with a worked application or study-specific design example. Five studies described trial outcomes, trial designs, or a deployed public-health testing system. Studies addressed adaptivity through changes to allocation designs, inferential procedures, or both, supported mainly by theoretical and simulation evidence. Among studies for which each metric was relevant, statistical power was numerically evaluated in 26 of 32 studies, type I error in 23 of 32, estimator bias in eight of 35, and interval coverage in five of 12. These findings indicate that numerical evaluation in the included studies has emphasized hypothesis testing properties more than estimation and uncertainty quantification. Adaptive allocation and statistical inference should therefore be evaluated jointly, with attention not only to type I error and power but also to estimator bias, precision, and interval coverage.}


\makeatletter
\let\ps@opening\ps@plain
\makeatother
\pagestyle{plain}
\maketitle

\section{Introduction}\label{sec:intro}

Response-adaptive allocation designs, in which treatment allocation probabilities are updated using accumulated data, have a long history in statistics \citep{thompson1933likelihood,robbins1952some}. Throughout this review, we use ``adaptive experiments'' and ``adaptive experimental designs'' for experiments using response-adaptive or related bandit-based allocation, and ``adaptive allocation'' for the process of assigning treatments and updating allocation probabilities using accumulating data. Contemporary examples include response-adaptive randomization and covariate-adjusted response-adaptive (CARA) designs in clinical trials, and contextual bandit algorithms in mobile-health systems. These designs can allocate more participants to promising treatments (e.g., play-the-winner rules \citep{hu2006theory}), reduce exposure to inferior treatments \citep{villar2015multiarmed}, or use real-time participant information and accumulated outcomes to adapt intervention choices while balancing exploration and exploitation, as in contextual bandit methods \citep{tewari2017ads}.

Adaptive allocation can alter the sampling distribution of treatment-effect estimators and test statistics relative to that arising under conventional randomization \citep{shin_are_2019}. Under response-adaptive randomization, treatment allocation probabilities depend on previously observed outcomes and may also depend on baseline covariates, interim estimates, or posterior quantities, depending on the design. Consequently, inferential procedures developed under non-adaptive allocation need not retain their nominal properties when applied without modification. Depending on the allocation mechanism and inferential procedure, adaptivity may affect estimator bias and precision, confidence interval coverage, type I error, and statistical power \citep{hadad2021confidence,fda2019adaptive}. Establishing when conventional procedures remain valid, when adjustment is required, and which inferential procedures are appropriate for a given adaptive design is therefore an important methodological and practical problem.

Recent methodological work has addressed the inferential challenges of adaptive allocation through approaches operating at both the design and analysis stages. Design-stage approaches include selecting or constraining target allocations to balance statistical power and participant benefit \citep{antognini2019optimal,frieri2021optimal}, whereas analysis-stage approaches include variance-stabilized bootstrap methods \citep{antognini2022vsb}, inverse probability weighted and semi-parametric estimators \citep{hadad2021confidence,zhu2023covariate,mukherjee2025covariate}, and targeted maximum likelihood estimation \citep{chambaz2014inference}. Bayesian approaches have also been used to guide adaptive allocation and support inferential decision making \citep{houston2020attacc,holmhansen2017development}. These approaches address distinct, although related, statistical objectives. Some studies seek to construct allocation algorithms with desirable operating characteristics \citep{antognini2019optimal,frieri2021optimal}, whereas others seek valid procedures for estimation, uncertainty quantification, or hypothesis testing under adaptive allocation  \citep{chambaz2014inference,hadad2021confidence}. Papers reporting trial findings may, in turn, describe the adaptive allocation procedure and subsequent analyses while providing limited detail on how the analyses account for adaptive allocation \citep{sasaki2022lowdose}. 

The review by \citet{robertson_response-adaptive_2023} examined methodological and practical debates surrounding response-adaptive randomization in clinical trials, including approaches to valid inference. It highlighted the relatively limited reporting of estimation metrics in evaluations of these designs, alongside the need for valid finite-sample inferential procedures with less loss of power. Building on these observations, we conducted a systematic review to examine how inferential procedures have been developed and evaluated under adaptive allocation in medical, health, and biostatistical research.

We were guided by PRISMA 2020 in reporting study identification, screening, and synthesis \citep{page2021prisma}, with PRISMA-S used to guide reporting of the search strategy \citep{rethlefsen2021prismas}. Because the included studies differ substantially in their adaptive designs, estimands, outcome types, inferential procedures, and analytical characteristics, we use a structured methodological evidence map, presented through study-level tables and summary counts organized around the four review questions, together with narrative comparisons. We characterize the adaptive designs that have been studied, the inferential procedures considered, the methods used to justify or evaluate inference, and the evidence reported for type I error, confidence interval coverage, estimator bias, and statistical power. Our objective was to identify patterns and gaps in how inferential validity has been established and evaluated across adaptive experimental designs.

\section{Scope and review questions}\label{sec:scope}

This review focuses on statistical inference under adaptive experimental designs in which treatment allocation probabilities are updated using accumulated data. We consider response-adaptive randomization, CARA designs, and bandit-based adaptive algorithms that are relevant to clinical trials, mobile health, public health, and related areas of biostatistical research. Our primary interest is inference for treatment effects following adaptive allocation, including estimation,  uncertainty quantification, and hypothesis testing, rather than the optimization or computational implementation of the allocation algorithms themselves. 

The review is organized around four primary questions:

\begin{enumerate}
\item[Q1.] Which adaptive experimental designs have been studied, and what scientific and statistical objectives motivate their use?
\item[Q2.] What inferential procedures are used to analyze data from adaptive experimental designs?
\item[Q3.] How have studies addressed and justified statistical inference under adaptive allocation through the allocation design, statistical analysis, or both, and what intended targets and supporting evidence were reported?
\item[Q4.] How has inferential performance been evaluated and reported, with a focus on statistical power,  type I error, estimator bias, interval coverage, and relevant design-inference trade-offs?
\end{enumerate}

We describe study characteristics, including adaptive design, outcome type, sample size, number of treatments, and application domain, to provide context for the synthesis of inferential methods and reported performance. The following sections describe the search strategy, inclusion and exclusion criteria, the extraction process, and the synthesis methods used in this review.

\section{Methods}\label{sec:methods}

\subsection{Protocol and registration}

Reporting of the review was guided by PRISMA 2020 \citep{page2021prisma}. The PRISMA 2020 checklist and the PRISMA 2020 for Abstracts checklist are provided in Supplementary Tables~\ref*{tab:prismachecklist} and~\ref*{tab:prismaabstractchecklist}, respectively. An internal, unpublished protocol draft was prepared during the review. The review was not registered.  Eligibility  criteria, study selection, data extraction, and synthesis procedures are described below. One eligibility criterion concerning publication status was refined after screening had begun but before full-text review. The timing and rationale for this refinement are described in the Eligibility criteria subsection.

\begin{figure*}[htbp]
\centering
\includegraphics[width=0.86\textwidth]{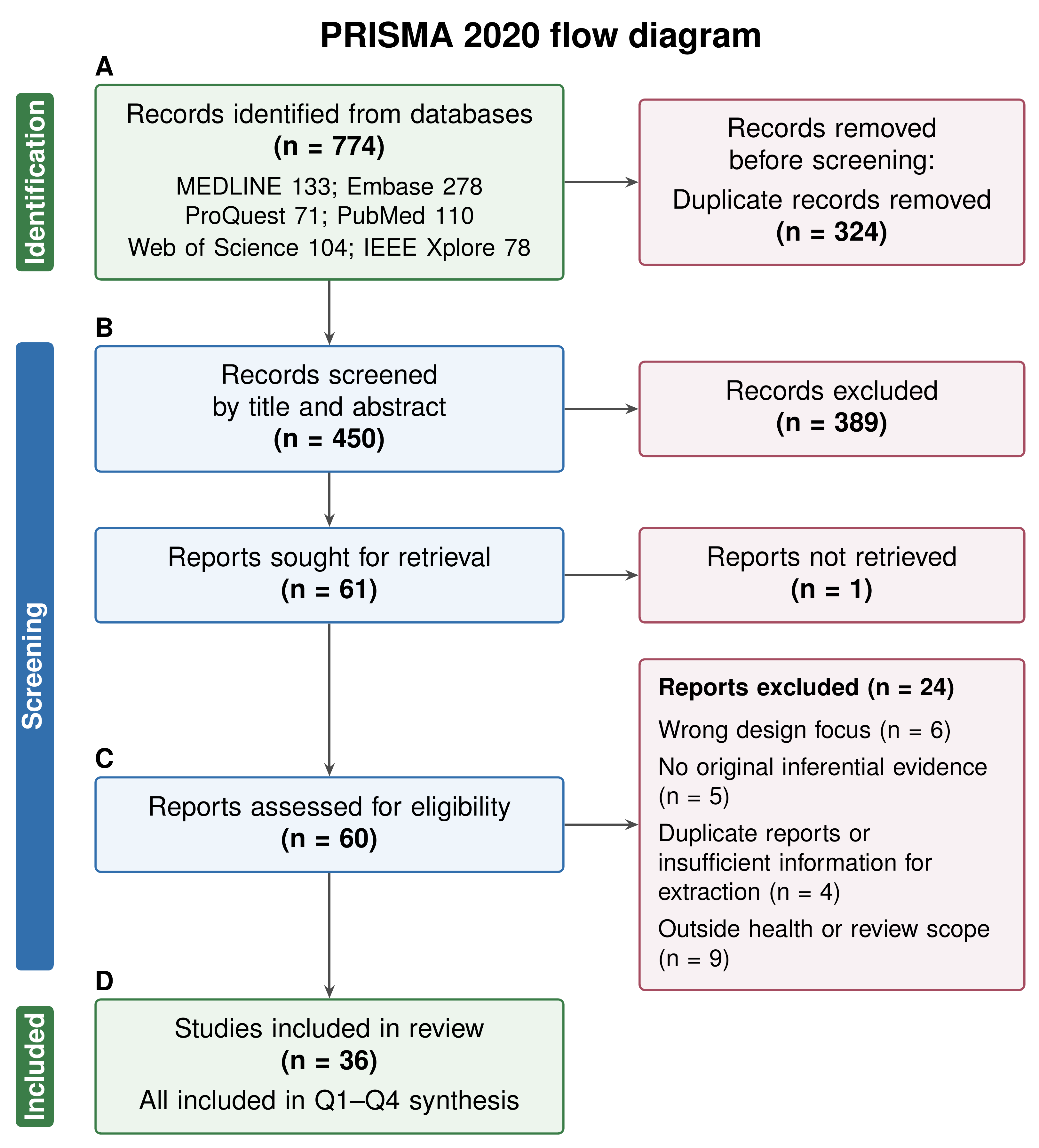}
\caption{PRISMA 2020 flow diagram for study selection. Of 774 records identified, 324 duplicates were removed and 450 records were screened. After title-and-abstract screening, 389 records were excluded and 61 records were retained for full-text retrieval. One full text could not be obtained; 60 were assessed for eligibility and 24 were excluded, leaving 36 studies included in the synthesis. Labels A--D provide reference points for the Methods section.}
\label{fig:prisma}
\end{figure*}

Figure~\ref{fig:prisma} summarizes the search, identification, and selection of studies. Labels A--D provide reference points for the methods described below.
 
\subsection{Eligibility criteria} \label{subsec:eligibility}

Studies were eligible if they investigated adaptive experimental designs, including CARA, multi-armed bandit, and contextual bandit designs. Eligible studies were additionally required to address statistical inference following adaptive allocation by developing, applying, or evaluating an estimation or hypothesis testing procedure; to evaluate one or more inferential characteristics such as bias, statistical power, type I error, or confidence interval coverage; or to provide methodological,  theoretical, simulation-based, or applied evidence relevant to inferential validity. Studies were restricted to those published in English from January 2010 through October 2025 for which an accessible full text provided sufficient information for data extraction.

Studies were excluded if they did not involve adaptive allocation or bandit-based adaptive design, or if they described or optimized an adaptive algorithm without addressing statistical inference following adaptive allocation. We also excluded studies focused on sample-size re-estimation (interim or post-study), treatment selection, or historical borrowing when they did not involve adaptive allocation; dose-finding studies without adaptive allocation; and studies of reinforcement learning methods (e.g., Q-learning) that did not address inference for an experimental treatment comparison. Duplicate papers were excluded. Abstracts, presentations, and protocols were excluded when they lacked sufficient methodological or empirical information for extraction. Studies outside health, biomedical, or related biostatistical settings and publications not in English were also excluded.

For dose-finding and bandit studies specifically, studies were eligible when treatment or intervention allocation responded to accruing outcomes and the study evaluated statistical inference for the resulting comparison. Studies concerned only with optimizing a dose, allocation algorithm, or control policy without corresponding inferential evaluation were excluded. Because the initial search retrieved a large number of non-peer-reviewed records, eligibility was subsequently restricted to peer-reviewed journal articles and conference studies presenting original methodological or empirical research, including studies reported in conference abstracts with sufficient information for extraction. This criterion was introduced after initial screening but before full-text review.

\subsection{Information sources and search}

All six sources were searched on 13 October 2025 for records published from January 2010 through October 2025. Searches were conducted on Ovid MEDLINE, Embase, ProQuest, PubMed, Web of Science, and IEEE Xplore (Figure~\ref{fig:prisma}, A). The core search strategy combined four concepts: adaptive experimentation or bandit methods, statistical inference, hypothesis testing, and application context. Terms within each concept were combined with \texttt{OR}, and the four concept blocks were combined with \texttt{AND} (Table~\ref{tab:searchstrategy}). The search yielded 774 records before duplicate removal: 133 from MEDLINE, 278 from Embase, 71 from ProQuest, 110 from PubMed, 104 from Web of Science, and 78 from IEEE Xplore. IEEE Xplore used a shortened query because of interface constraints. The shared concept-level search strategy is provided in Supplementary Section~\ref*{app:search}.

\begin{table}[!htbp]
\caption{Concept-level search strategy used to identify records. An asterisk (*) denotes truncation to match words beginning with the same stem (e.g., \texttt{trial*} matches trial and trials).}
\label{tab:searchstrategy}
\centering
\sffamily\fontsize{8}{10}\selectfont
\setlength{\tabcolsep}{3pt}
\setlength{\arrayrulewidth}{0.6pt}
\renewcommand{\arraystretch}{1.3}
\begin{tabularx}{\linewidth}{|L{0.115\linewidth}|L{0.235\linewidth}|L{0.08\linewidth}|Y|}
\hline
\textbf{Concept 1} &
Adaptive\newline Experimentation /\newline Bandit Methods &
Search &
``Adaptive experiment*'' OR ``adaptive design*'' OR
``response-adaptive randomization'' OR ``multi-armed bandit*'' OR
``bandit algorithm*'' OR ``adaptive allocation'' OR ``adaptive trial*'' OR
``dynamic treatment regime*'' OR ``reinforcement learning'' \\
\hline
& \multicolumn{3}{c|}{AND} \\
\hline
\textbf{Concept 2} &
Inference / Statistical\newline Evaluation &
Search &
``Statistical inference'' OR ``estimation'' OR ``bias correction'' OR
``Type I error'' OR ``coverage probability'' OR ``power analysis'' OR
``posterior estimation'' OR ``frequentist inference'' OR
``Bayesian inference'' OR ``False Positive Rate'' OR
``False Negative Rate'' OR ``confidence interval*'' \\
\hline
& \multicolumn{3}{c|}{AND} \\
\hline
\textbf{Concept 3} &
Hypothesis Testing /\newline Evaluation &
Search &
``Hypothesis test*'' OR ``statistical test*'' OR ``null hypothesis'' OR
``p-value'' OR ``test statistic'' OR ``sequential testing'' OR
``stopping rule*'' OR ``error control'' OR ``Bayes Factor'' OR
``Bayesian Analysis'' OR ``Posterior Probability of Arm Being Optimal'' \\
\hline
& \multicolumn{3}{c|}{AND} \\
\hline
\textbf{Concept 4} &
Context / Application\newline Domain &
Search &
``Biostatistics'' OR ``medical trial*'' OR ``clinical trial*'' OR
``health study'' OR ``learning experiment'' OR ``digital intervention'' OR
``behavioral experiment'' \\
\hline
\end{tabularx}
\end{table}

\subsection{Selection process}

Duplicates were identified and removed using Rayyan \citep{ouzzani2016rayyan}. After duplicate removal, 450 records were screened by title and abstract (Figure~\ref{fig:prisma}, B) in two passes: the first removed records that did not concern adaptive experiments, and the second removed records that lacked original methodological, simulation, theoretical, or empirical evidence bearing on inferential performance. Reviewer~1 screened titles and abstracts and assessed full-text eligibility using the specified criteria. At both stages, uncertainty about eligibility or other concerns prompted review by Reviewer~3, who had access to Reviewer~1's assessment and rationale. Disagreements were resolved through discussion to reach a final eligibility decision. Screening prioritized sensitivity, and records for which the available information was insufficient to justify exclusion were discussed and retained for full-text review when uncertainty remained.

Sixty-one records were retained for full-text retrieval; one conference talk could not be obtained in full text, leaving 60 full texts assessed for eligibility (Figure~\ref{fig:prisma}, C). Twenty-four were excluded after full-text assessment: six for the wrong design focus, five for lack of original inferential evidence, four because they were duplicate papers or provided insufficient information for extraction, and nine because they fell outside the health or review scope (see Supplementary Table~\ref*{tab:appexcl} for details). Thirty-six studies met the inclusion criteria and were included in the synthesis (Figure~\ref{fig:prisma}, D). 

\subsection{Classification framework}

To answer Q1--Q4, we organized the terminology across the included studies, using a common framework to distinguish features of the adaptive design. We distinguished allocation algorithms, estimands, inferential procedures, and approaches used to address the inferential consequences of adaptive allocation.

The allocation algorithm in Q1 described how treatments were allocated or how such allocation probabilities were updated using accruing data. The estimand was the population-level quantity targeted by the analysis. For example, in a hypothetical trial comparing a vaccine with placebo, the estimand could be the difference in six-month infection risk under assignment to the vaccine versus placebo in the population eligible for the trial. These were recorded separately because the allocation algorithm did not determine the inferential target or the analysis used to estimate it.

The inferential procedures, investigated in Q2, described how studies estimated an estimand, quantified uncertainty, or tested a hypothesis. Examples included estimation methods, confidence interval procedures, frequentist hypothesis tests, or Bayesian decision rules. Q2 therefore identified the inferential procedures used and broadly classified final analyses, whereas Q3 examined how studies addressed and justified inference under adaptive allocation through the allocation design, the statistical analysis, or both.

We also distinguished the objectives in Q3 from their reported performance. Q3 described the studies' objectives and evaluation targets, such as power, estimation precision, or participant benefit, and whether these approaches were supported by theory, simulation or application. Q4 examined how inferential performance was evaluated and reported, with a focus on statistical power and type I error for hypothesis testing, estimator bias for point estimation, and interval coverage for uncertainty quantification. Coverage describes how often confidence intervals contain the true value of the estimand across repeated datasets. For each evaluation metric, we recorded whether it was relevant, mentioned, and measured using the definitions in the notes to Table~\ref{tab:q4}. Q4 also considered the reported performance findings and trade-offs between inferential objectives and participant benefit.

\subsection{Data extraction}

Data from the included studies (Figure~\ref{fig:prisma}, D) were extracted into a structured Google Sheets workbook with sheets aligned to the four review questions: adaptive design characteristics and objectives (Q1), reported inferential procedures (Q2), adjustment or correction approaches (Q3), and reported performance metrics and trade-offs (Q4). Extracted fields included study identifier (number), citation, application domain, study type, number of treatment arms, outcome type, adaptive target or allocation algorithm, comparator, sample size, estimand or objective, inference method, justification of inference under adaptive allocation, reported $p$-values or intervals, theoretical and simulation evidence, real-data application, type I error, confidence-interval coverage, estimator bias, statistical power, and inferential trade-offs.

For papers reporting several analyses or simulation scenarios, Reviewer~1 selected findings for initial extraction case by case, usually examining comparison tables first and then consulting the Results and Methods sections. Entries were coded as NR (not reported) when relevant information was not reported, and NA (not applicable) when the item did not apply to the design or analysis. The cleaned study-level Q1--Q4 tables are provided in the Study-Level Extraction supplement (Section~\ref*{app:full-extraction}).

\subsection{Data extraction verification}\label{subsec:extraction-verification}

Three reviewers participated in data extraction and verification. The reporting workbook contains field-specific coding guidance, including column-specific definitions, the distinction between NR and NA, and instructions to code information from the source articles. Reviewer~1 extracted the Q1--Q4 fields for each included study. Reviewer~2 then reviewed Reviewer~1's entries against the source articles and the same coding guidance. After identifying discrepancies while reviewing Reviewer~2's comments, Reviewer~1 consulted AI tools for feedback on possible resolutions. Reviewer~1 evaluated the feedback, corrected entries where appropriate, and referred unresolved disagreements to Reviewer~3, who reviewed the source article and the extracted entries and made the final decisions. The reported quantitative summaries were calculated from the updated reporting workbook.

\subsection{Synthesis methods}

We summarized the extracted data using study-level counts and structured narrative comparisons organized according to the four review questions. This approach was chosen to compare how inference was conducted and evaluated across studies with different estimands, adaptive designs, and evaluation settings. Summary tables described the distribution of adaptive design characteristics and objectives, inferential procedures, estimands, correction methods, and reported performance metrics. Comparisons across studies were descriptive and were used to identify recurring methodological approaches, patterns in the evaluation of inferential properties, and gaps in reporting. 

During verification and synthesis, we checked the type of evidence supporting the reported findings, distinguishing theoretical justification from simulation and applied results. We considered this evidence in relation to the reported assumptions and evaluation settings, and qualified interpretations where support was limited or unclear.

\subsection{Topic-modeling description of the screening records}

As a secondary descriptive analysis, we applied topic modeling to the title, abstract, and keyword text of the 450 screened records after duplicate removal (Figure~\ref{fig:prisma}, B). Topic assignments did not determine eligibility. 
Each record was represented by a row in a bag-of-words document-term matrix restricted to the 1,000 most frequent terms, with terms retained if they appeared in at least two records and in no more than 95\% of records.  We fit a latent Dirichlet allocation model \citep{blei2003latent} using the scikit-learn package~\citep{pedregosa2011scikit}.  

We compared paper-level models with three to eight topics and sentence-level models with three to six topics. We examined in-sample perplexity \citep{blei2003latent}, mean entropy of document-topic weights \citep{shannon_info}, and the median difference between the two largest topic weights. These measures served as diagnostics of model fit and topic concentration rather than as automated selection rules. The number of topics was selected based on interpretability considered alongside these diagnostics. See Supplementary Section~\ref*{app:topicmodel} for details.

\FloatBarrier
\section{Results}\label{sec:results}

\subsection{Overview of the evidence (Q1)}

The evidence base was primarily methodological, with relatively few studies reporting implementation in clinical trials. Among the 36 included studies, 25 focused on developing or evaluating adaptive designs or inferential methods, primarily using theoretical arguments and simulation \citep[e.g.,][]{baldiantognini2022simple, baas2025exact}; six included a worked application or study-specific design example \citep[e.g.,][]{frieri2021optimal,sverdlov2014efficient}, and five were applied studies \citep{guglielmetti2025oral, houston2020attacc, bernhardt2023avertdose, sasaki2020longterm, bastani2022interpretable}. Of these five studies, two presented trial outcomes, two described trial designs, and one described a deployed public-health testing system.

The application setting was also concentrated in clinical research. Thirty-three studies focused on clinical trials or dose-response settings \citep[e.g.,][]{guglielmetti2025oral, dette2013efficiency}, two considered mobile-health adaptive interventions or policy evaluation \citep{huch2024robust, liao2021offpolicy}, and one addressed public health testing allocation \citep{ bastani2022interpretable}. 
Table~\ref{tab:q1} summarizes these study characteristics. The papers contributing to each count are listed in Supplementary Table~\ref*{tab:papersq1}, and study-level details are provided in Study-Level Extraction Tables~\ref*{tab:appq1}--\ref*{tab:appq1b2}.
 
Taken  together, the  evidence is concentrated on the development and evaluation of methods, with fewer studies describing their implementation in specific trials or deployed systems. Conclusions about inferential properties therefore depend mainly on the assumptions and scenarios examined in theoretical and simulation studies. Based on the extracted results for each study, eight studies combined Bayesian allocation with decision rules evaluated using frequentist criteria such as type I error or power, while three combined Bayesian allocation with a frequentist final analysis. Here, frequentist evaluation refers to assessing the type I error or power of a procedure for comparing treatments across repeated trials, even when conclusions about treatment differences are based on Bayesian posterior probabilities. A frequentist final analysis, in contrast, uses measures such as $p$-values or confidence intervals to draw conclusions from the observed trial data. The allocation algorithm itself does not determine the inferential procedure used to estimate or test treatment effects, highlighting the need to describe the allocation and inferential procedures separately. Mobile-health and public health adaptive-allocation settings are less represented, so there is less direct evidence on inference in these settings.

\begin{table*}[!htbp]
\footnotesize 
\setlength{\tabcolsep}{4pt}
\caption{Q1. Design and algorithm overview across the 36 included studies. Outcome categories and allocation algorithm families are not mutually exclusive. Inference groups reflect the main framework recorded for each study.\label{tab:q1}}
\centering
\begin{tabularx}{\textwidth}{L{0.20\textwidth}L{0.30\textwidth}L{0.05\textwidth}Y}
\toprule
Dimension & Category & $n$ & Interpretation \\
\midrule
Evidence type & Methodological & 25 & Primary evidence is mostly methods and design work. \\
 & Methodological with applied illustration & 6 & Includes a worked application or study-specific design example. \\
 & Applied study (trial or deployed system) & 5 & Applied evidence is present but sparse. \\
\midrule
Application setting & Clinical trial or dose-response & 33 & Strongly aligned with health-methods scope. \\
 & Mobile-health contextual bandit & 2 & Broadens scope beyond conventional trials. \\
 & Public-health testing allocation & 1 & Non-trial allocation problems with inference. \\
\midrule
Number of Treatments & Two-arm ($K=2$) & 15 & Common for focused inference questions. \\
 & General or lower-bound multi-arm ($K\geq 2$ or $K\geq 5$) & 16 & Method stated for a family of arm counts; includes one study with both two-arm and $K\geq 2$ settings. \\
 & Fixed multi-arm example ($K=3$--$6$) & 5 & Supports multiplicity and best-arm discussion. \\
\midrule
Outcome type & Binary & 21 & Central to response-adaptive operating-characteristic work. \\
 & Continuous & 18 & Supports Wald, likelihood, and power calculations. \\
 & Time-to-event & 8 & Raises censoring and delayed-response issues. \\
 & Count & 4 & Handled by several general methods. \\
 & Ordinal or composite & 2 & Broadens relevance, complicates comparison. \\
\midrule
Allocation algorithm family & Optimal or constrained allocation & 22 & Treats the allocation algorithm as part of inferential design. \\
 & Bayesian / posterior-probability RAR & 11 & Posterior information guides treatment allocation. \\
 & Covariate-adjusted RAR (CARA) & 6 & Covariates enter the allocation mechanism. \\
 & Urn, play-the-winner, or sequential-MLE & 4 & Includes classical allocation algorithms and their associated allocation targets. \\
 & Bandit or Gittins-index & 3 & Bandit-driven allocation, mostly outside trials. \\
 & Two-stage or group-sequential RAR & 11 & Combines adaptive allocation with staged design changes or group-sequential testing. \\
 & Doubly adaptive biased coin design & 3 & Implements a target allocation. \\
 \midrule
 Inference and evaluation & Other studies using frequentist inference & 22 & Includes methods for estimation and testing, such as likelihood-based estimation, bootstrap procedures, and Wald or exact tests.\\
 & Bayesian allocation with frequentist evaluation of decision rules & 8 & Evaluate how often treatment-comparison rules give false-positive conclusions or detect true treatment differences across repeated trials. \\
 & Bayesian allocation with a frequentist final analysis & 3 & Treatment conclusions use hypothesis tests or confidence intervals after Bayesian allocation. \\
 & Bandit or policy-evaluation studies targeting reward, policy value, or prevalence & 3 & Use reward or policy-value estimates to assess interventions, and estimates of infection prevalence to help decide which passengers receive COVID-19 tests. \\
\bottomrule
\end{tabularx}
\end{table*}

\FloatBarrier
\subsection{How inference is conducted and reported (Q2)}

Most studies defined or carried out some form of formal hypothesis test or decision rule. Among the 36 included studies, 32 included a formal test or decision rule, while the remaining four reported estimates without a formal treatment-comparison test. With respect to how final analysis was conducted, 23 studies were classified as adjusted. Another 12 studies relied primarily on properties of the adaptive design to justify inference. In contrast, one applied study used a conventional analysis without explicit adjustment for adaptive allocation algorithms.

In terms of inferential procedures, Bayesian or posterior-probability procedures were the most common, appearing in nine studies \citep[e.g.,][]{houston2020attacc, park2022personalized}. These were followed by Wald or score tests \citep[e.g.,][]{mukherjee2025covariate, magirr2011block}, with seven studies; other likelihood-based methods \citep[e.g.,][]{baldiantognini2022simple, dette2013efficiency}, with three studies; exact or conditional testing \citep[e.g.,][]{baas2025exact}, with three studies; inference and evaluation for bandit-based designs \citep[e.g.,][]{huch2024robust}, with three studies; targeted maximum likelihood or semi-parametric estimation with three studies \citep[e.g.,][]{mukherjee2025covariate}; off-policy value estimation \citep[e.g.,][]{ liao2021offpolicy} with two studies; and the variance-stabilized bootstrap \citep{antognini2022vsb} with one study. These categories are not mutually exclusive: five studies contributed to two categories, while ten did not map to one of the named categories summarized in Table~\ref{tab:q2}. Further, the eight categories did not cover all procedures recorded in the extraction. For example, \citet{chang2013sequential} proposed stopping data collection when treatment effects could be estimated with a specified precision. The papers contributing to each count are listed in Supplementary Table~\ref*{tab:papersq2}, and study-level details are provided in Study-Level Extraction Tables~\ref*{tab:appq2}--\ref*{tab:appq2b}.

\begin{table*}[!htbp]
\footnotesize
\caption{Q2. Inferential procedures and analysis across the 36 included studies. The eight categories in the final block may overlap: five studies contributed to two categories and ten to none, so these counts need not sum to 36. TMLE was counted separately from other likelihood-based methods unless a distinct likelihood-based analysis was also reported.\label{tab:q2}}
\centering
\begin{tabularx}{\textwidth}{L{0.26\textwidth}L{0.05\textwidth}Y}
\toprule
Inferential Procedure & $n$ & Interpretation \\
\midrule
Formal test or decision rule reported & 32 & Formal testing or decision-making is more common than estimation alone.  \\
Estimation or value only (no formal test) & 4 & Type I error is not the primary reported quantity for these studies. \\
\midrule
Adjusted final analysis & 23 & Most studies do not treat standard fixed-trial inference as automatically valid. \\
Design-based final analysis & 12 & Inferential performance and validity are analyzed through properties of the adaptive design. \\
Conventional final analyses (coded as NA) & 1 &
One applied study analyzes long-term outcomes conventionally; no adaptation-specific correction is reported. \\
\midrule
Bayesian or posterior-probability family & 9 & Some studies assess posterior decision rules using frequentist type I error and power criteria. \\
Wald or score tests & 7 & Familiar but sensitive to small samples and allocation imbalance. \\
Other likelihood-based methods & 3 & Asymptotic or target-allocation arguments. \\
Exact or conditional testing & 3 & Type I error control is established under the stated model and design assumptions. \\
Inference and evaluation for bandit-based designs & 3 &  Bandit-based designs address several inferential targets, including policy value, prevalence, and treatment comparisons. \\
TMLE or semi-parametric estimation & 3 & Used with CARA designs in all three studies. \\
Off-policy value estimation & 2 & Estimates outcomes under one allocation algorithm using data collected under another. \\
Variance-stabilized bootstrap & 1 & A general adjusted route across outcome types. \\
\bottomrule
\end{tabularx}
\end{table*}

Two reporting patterns are worth noting here. First, evidence supporting inferential validity was predominantly theoretical or simulation-based rather than based on results of applied trials. Thus, the literature generally established characteristics of inferential procedures under specified assumptions and data generating scenarios rather than through evidence from applied implementation. Second, the four studies without a formal treatment-comparison test reported estimates or intervals. Type I error was not the primary reported quantity for these studies.

\FloatBarrier
\subsection{Corrections and alternative inferential procedures under response-adaptivity (Q3)}

The third question examined how studies addressed the inferential consequences of adaptive allocation. Overall, the literature did not converge on a single correction or adjustment method. Instead, studies addressed inferential properties at the treatment allocation stage and at the data analysis stage.

At the treatment allocation stage, studies examined how adaptive design choices affected the efficiency of estimation or modified the allocation algorithm to improve characteristics of inferential procedures. Examples included constrained-optimal or rescaled allocation targets \citep{antognini2019optimal, baldiantognini2022simple}, model-based and doubly-adaptive biased-coin allocation \citep{liu2025modelbased, zhu2017statistical}, and initial allocation procedures used to obtain the parameter estimates needed to begin the main adaptive allocation procedure \citep{haines2015startup}.

At the data analysis stage, approaches focused on estimation and testing procedures that accounted for adaptive allocation. These included targeted maximum likelihood and semi-parametric estimators \citep{chambaz2014inference, zhu2023covariate}, exact or conditional finite-sample tests \citep{baas2025exact}, familywise error-control procedures based on conditional invariance \citep{robertson2019familywise, gutjahr2011familywise}, variance-stabilized bootstrap methods \citep{antognini2022vsb}, and off-policy value estimators~\citep{liao2021offpolicy}.

While Q2 identified which inferential procedures studies used, Q3 examined how the allocation design or analysis accounted for adaptivity. It also examined what these approaches aimed to achieve and what evidence was reported to support them. For example, a study using posterior-probability decision rules contributed to the Bayesian or posterior-probability category in Q2 \citep[e.g.,][]{park2022personalized}. In Q3, the same study illustrated how simulations were used to select decision thresholds with the goal of meeting specific type I error and power targets. In practice, some studies combined both allocation-stage and analysis-stage approaches: the allocation algorithm was designed with the final analysis in mind, and the analysis was then justified for that specific algorithm \citep[e.g.,][]{chambaz2014inference,zhu2023covariate}. For this reason, the extraction did not support mutually exclusive counts of allocation-stage and analysis-stage approaches. 

The evidence reported in the included studies was predominantly theoretical and simulation-based. Among the 36 studies, 28 provided  some form of theoretical justification, such as a derivation, asymptotic argument, or theorem. Thirty-two studies evaluated inferential or design operating characteristics through simulations. Ten studies included applications to real data or a real trial. These forms of evidence could overlap within a study (Study-Level Extraction Table~\ref*{tab:appq3b}). 

In terms of intended targets, the methods most often focused on statistical power, which appeared in 27 studies, and type I error, which appeared in 24 studies. These were followed by robustness in 18 studies, ethical or participant-benefit-oriented allocation in 18 studies, estimation precision in 13 studies, sample-size efficiency in 11 studies, bias in eight studies, confidence interval coverage in five studies and direct patient benefit in five studies. These categories were non-mutually exclusive and included inferential characteristics, design objectives, and design operating characteristics. Table~\ref{tab:q3} summarizes these objectives and evaluation targets. The papers contributing to each count are listed in Supplementary Table~\ref*{tab:papersq3}, and study-level details are provided in Study-Level Extraction Table~\ref*{tab:appq3a2}.

\begin{table*}[!htbp]
\footnotesize
\caption{Q3. Reported objectives and evaluation targets across the 36 included studies. Categories are not mutually exclusive; a study can contribute to multiple rows.\label{tab:q3}}
\centering
\begin{tabularx}{\textwidth}{L{0.26\textwidth}L{0.10\textwidth}Y}
\toprule
Target & Number of studies, $n$ & Interpretation \\
\midrule
Statistical power & 27 & Most frequently considered inferential quantity. \\
Type I error & 24 & Second most frequently considered inferential quantity. \\
Robustness & 18 & Evaluation under misspecification, time trends, or design details. \\
Ethical / participant-benefit allocation & 18 & Participant benefit incorporated as a design objective. \\
Estimation precision & 13 & Standard errors, information, or efficiency. \\
Sample-size efficiency & 11 & Average sample size or stopping. \\
Estimator bias & 8 & Finite-sample or selection-induced bias. \\
Confidence-interval coverage & 5 & Interval validity, less often considered. \\
Direct patient benefit & 5 & Failures avoided or responders gained. \\
\bottomrule
\end{tabularx}
\end{table*}

\FloatBarrier
\subsection{Reported performance and trade-offs (Q4)}

Whereas Q3 examined how studies addressed adaptivity and what they aimed to achieve, Q4 examined whether relevant inferential performance metrics were evaluated and reported, and what these evaluations showed. We focused on four core metrics used to evaluate inferential procedures: statistical power and type I error for hypothesis testing, estimator bias for point estimation, and interval coverage for uncertainty quantification. For each, we considered whether it was relevant to the study's inferential objectives and whether it was mentioned and quantitatively measured in the study (Table~\ref{tab:q4}).

A metric was considered relevant when it was meaningful for the inferential procedure addressed in the study, regardless of whether the authors explicitly mentioned or measured it. For example, estimator bias was considered relevant to a study that estimated a treatment effect, even if the study did not evaluate whether the estimator overestimated or underestimated the effect on average. Among the 36 included studies, power and type I error were each relevant to 32 studies, estimator bias to 35, and  interval coverage to 12.

An evaluation metric was counted as mentioned when the paper explicitly discussed, targeted, or evaluated the metric for its own method or design. Among the included studies, power was mentioned in 30 studies, and type I error in 29. In comparison, 20 studies mentioned estimator bias and six mentioned confidence interval coverage. 

Lastly, a metric was counted as measured when a numerical evaluation was reported in the text, a table, or a result figure. A stated nominal significance level, a planned power target, or a reported confidence interval on its own did not qualify as a numerical performance evaluation. Among studies for which each metric was relevant, power was measured in 26 of 32 studies (81.3\%), type I error in 23 of 32 (71.9\%), estimator bias in eight of 35 (22.9\%), and confidence interval coverage in five of 12 (41.7\%). Thus, numerical evaluation was more common for hypothesis testing properties than for estimator bias or interval coverage, even after accounting for differences in the applicability of each metric.

Further, within-study simulation comparisons also illustrated why allocation objectives, such as participant benefit, and inferential performance should be evaluated jointly: improvements in one did not necessarily correspond to improvements in the other. For example, in a simulated two-arm trial with 600 participants and a bounded continuous outcome, a CARA design combined with targeted maximum likelihood estimation achieved higher power than complete randomization with a t-test (85.0\% versus 74.2\%) while also allocating a greater proportion of participants to the better performing treatment (58.4\% versus 50.0\%). Type I error was similar under the two approaches (5.24\% and 5.18\%, respectively, at a nominal significance level of 5.0\%) \citep{zhu2023covariate}. Thus, in this setting, the adaptive design and corresponding analysis improved power while increasing allocation to the better performing treatment. 

In a different two-arm simulation setting with a normally distributed outcome and 116 participants, the Gittins index allocation algorithm assigned substantially more participants to the better performing treatment than equal randomization (87.9\% versus 50.0\%), but this greater allocation to the better treatment was accompanied by substantially lower power (23.7\% versus 90.0\%) \citep{smith2018bayesian}, using the testing threshold specified for each design. These examples showed that greater allocation to the better performing treatment did not necessarily imply higher power. 

The examples also showed why performance findings should be interpreted in the context of the allocation algorithm, inferential procedure, and evaluation setting. Greater allocation to the better-performing treatment or higher power on their own did not establish low estimator bias or adequate interval coverage. The papers contributing to each count are listed in Supplementary Table~\ref*{tab:papersq4}, and study-level findings are provided in Study-Level Extraction Tables~\ref*{tab:appq4}--\ref*{tab:appq4b2}.

\begin{table*}[!htbp]
\footnotesize
\caption{Q4. Reporting of metrics used to evaluate inferential
performance across the 36 included studies. Values are counts
of studies.\label{tab:q4}}
\centering
\renewcommand{\arraystretch}{1.15}

\begin{tabularx}{\textwidth}{
  @{}lrrr>{\raggedright\arraybackslash}X@{}
}
\toprule
Metric & Relevant & Mentioned & Measured & Reporting pattern \\
\midrule
Statistical power
& 32 & 30 & 26
& Most frequently measured, both overall and among studies
for which it was relevant. \\[4pt]

Type I error
& 32 & 29 & 23
& Widely mentioned, although fewer studies provided
numerical evaluations. \\[4pt]

Estimator bias
& 35 & 20 & 8
& Largest gap between relevance and numerical evaluation. \\[4pt]

Interval coverage
& 12 & 6 & 5
& Relevant to a smaller group of studies; fewer than half
provided numerical evaluations. \\
\bottomrule
\end{tabularx}

\medskip
\begin{minipage}{\textwidth}
\footnotesize
\textit{Notes:} Values are counts of studies.
\textbf{Relevant}: the metric meaningfully evaluates an
inferential objective or procedure addressed by the study.
\textbf{Mentioned}: the paper explicitly discussed, targeted,
or evaluated the metric for its own method or design.
\textbf{Measured}: the paper reported a numerical evaluation
of the metric under specified conditions.
Measured studies are also counted as mentioned and relevant.

\end{minipage}
\end{table*}

\FloatBarrier
\subsection{Topic distribution of the screening records }

As an exploratory analysis, we further examined how the included studies were distributed and clustered across the five topics identified among the 450 screened records. The largest topic did not have the highest proportion of records included in the review. As shown in Table~\ref{tab:topics}, the largest topic, ``adaptive and sequential trial design'', contained 215 records (47.8\% of the screening records), of which 14 were included in the data extraction (6.5\%). In comparison, the topic concerning ``optimal and adaptive treatment designs'' had the highest within-topic inclusion percentage, with 20\% of records included (17 out of 85 records). Taken together, these two topics accounted for 31 of the 36 included studies. The remaining five included studies were distributed across the other three topics, which shows that restricting screening to these two topics would have missed eligible studies.

\begin{table}[!htbp]
\caption{Topic-modeling description of the 450 screened records, separated by final inclusion status. Records are assigned to the topic of highest weight in the five-topic paper-level model, and Included (\%) is the percentage of a topic's records that were finally included. The cross-tabulation is descriptive and was computed after all eligibility decisions were complete; no record was included or excluded on the basis of its topic assignment.\label{tab:topics}}
\centering
\begin{tabularx}{\linewidth}{Yrrrr}
\toprule
Topic label & Total & Included & Not included & Included (\%) \\
\midrule
Optimal and adaptive treatment designs & 85 & 17 & 68 & 20.0 \\
Adaptive and sequential trial design & 215 & 14 & 201 & 6.5 \\
Randomized controlled drug/vaccine trials & 50 & 3 & 47 & 6.0 \\
Deep Learning and Reinforcement Learning for detection and networked sensing & 81 & 1 & 80 & 1.2 \\
Funded clinical trials in cancer treatment & 19 & 1 & 18 & 5.3 \\
\midrule
All records & 450 & 36 & 414 & 8.0 \\
\bottomrule
\end{tabularx}
\end{table}

\FloatBarrier
\section{Discussion}\label{sec:discussion}

In this review, we brought together evidence on adaptive allocation designs, their inferential objectives and procedures, and the evaluation of inferential performance. Across the four review questions, the synthesis linked the allocation algorithm and estimand with the analysis used, its justification, and the metrics evaluated. This connection is important when interpreting what the available literature supports for adaptive experimental designs, and when considering how to adapt an existing approach to a different experimental setting.

The evidence was predominantly methodological, with 25 of the 36 studies classified as methods or design papers. Ten studies reported applications to real data or trials, including some of these methodological papers. The evidence was concentrated in clinical-trial and dose-response settings, with relatively few studies addressing mobile health. Theory and simulation were central to examining inferential procedures under specified assumptions, using the reported evaluation metrics. Interpreting these findings therefore required attention to the particular allocation algorithm and data analysis method used.

The combinations of Bayesian allocation algorithms with frequentist final analyses, or with decision rules evaluated using frequentist criteria, showed that the allocation algorithm alone did not identify how treatment effects were estimated or tested. Studies addressed adaptivity through the allocation design, modifications to the inferential procedures, or both. These approaches offered different ways to justify inference, but each needed to be interpreted in relation to its own assumptions and intended use. For experiment designers and investigators considering an adaptive design, the relevant issue is whether that justification applies to the allocation algorithm and analysis they plan to use. 

At the same time, the evidence supporting these analyses also differed in the inferential properties examined. Q3 identified the intended targets of approaches used to address adaptivity, whereas Q4 examined which metrics were discussed or numerically evaluated (Study-Level Extraction Table~\ref*{tab:appq3a2} and Supplementary Table~\ref*{tab:papersq4}). For example, a study concerned only with point estimation might not require an assessment of power, but estimator bias would still be relevant even if the authors did not discuss or evaluate it. Differentiating whether a metric was relevant, mentioned, or measured therefore helped identify gaps in reporting without assuming all four metrics applied to every study. Among studies for which each metric was relevant, numerical evaluation was more common for statistical power and type I error than for estimator bias and interval coverage. 

The within-study comparisons in Q4 further showed why allocation algorithms and inferential properties should be considered together. In these comparisons, increased allocation to the better treatment was accompanied by higher power in one setting and lower power in another \citep{zhu2023covariate,smith2018bayesian}. These findings concerned different combinations of allocation algorithms and inferential procedures. They indicate that allocating more participants to the better treatment does not on its own establish improved inferential performance. Evidence about testing also does not establish estimation accuracy or interval coverage when those properties are relevant. 

\subsection{Relation to prior reviews}

A related review is that of \cite{robertson_response-adaptive_2023}, which our search retrieved and excluded at full-text screening as a background review. The paper examines methodological and practical debates around response-adaptive randomization in clinical trials. It discusses available approaches to valid inference, highlights the need for valid finite-sample methods with less loss of power, and notes the limited reporting of estimation metrics. Our review complements the discussion from this prior work in scope, method, and output. Our synthesis includes CARA designs, which were not a focus of that review, as well as studies of mobile-health interventions and public-health testing. Studies were identified through database searches and summarized using structured extraction, with reporting guided by PRISMA. Our study-level tables link each included study's allocation design and estimand with its inferential procedures, approaches used to address adaptivity, and reported performance (Study-Level Extraction supplement, Section~\ref*{app:full-extraction}). By distinguishing whether each performance metric was relevant, mentioned, or numerically evaluated, we quantify reporting gaps in the included literature while accounting for differences in the applicability of the metrics.

\subsection{Implications for practice and reporting}

These findings suggest that papers on adaptive experiments should make the connection between the allocation algorithm and other design features, inferential procedures, and evaluation metrics explicit. Naming an allocation algorithm or an inferential procedure could be insufficient to explain why their combination is appropriate for the setting. Authors should identify the estimand, describe how inference accounts for the allocation algorithm actually used, and state the assumptions and evidence supporting the analysis.

We suggest that simulation studies of adaptive experimental designs report the operating characteristics relevant to their inferential procedures: type I error and power for hypothesis tests or treatment-comparison decision rules, estimator bias for point estimation, and coverage for interval estimation. These should be reported alongside realized allocation proportions and, where relevant, measures of participant benefit or efficiency. When a metric is not applicable, the paper should state this and explain why. Bias and coverage assess estimation and interval performance, respectively, and were less often numerically evaluated than power and type I error among the included studies for which each metric was relevant.

Two further recommendations follow from the synthesis. Applied studies should state whether the final inference is adjusted, design-based, or conventional. Where it is adjusted, they should identify the method used. In each case, they should explain why the analysis is appropriate for the allocation algorithm used. Papers and protocols should specify each adaptive feature and explain how the inferential procedure accounts for their combination. For example, when a trial combines adaptive allocation with group-sequential testing, the analysis should account for outcome-dependent assignment and control type I error across interim and final analyses. These features were combined in the CARA design and TMLE-based group-sequential procedure of \citet{chambaz2014inference}. 

The topic modeling analysis also has an implication for identifying relevant literature in future reviews. Although eligible studies were concentrated in two topics, restricting screening to those topics would have omitted five of the 36 included studies (Table~\ref{tab:topics}). Eligibility further depended on the study's design and inferential content, which required assessment against the review criteria. These findings suggest caution about using topic assignments alone for automated exclusion.

\subsection{Limitations}

This review brings together theoretical, simulation-based, and applied evidence from studies of adaptive experiments with different estimands, assumptions, and objectives. For example, the clinical-trial-inspired simulation from \cite{zhu2023covariate} evaluated power and treatment-allocation proportions, while \cite{sasaki2020longterm} reported prognostic hazard ratios with confidence intervals from clinical trial data. These results provide different kinds of evidence, which limits direct comparisons of inferential performance across studies. 

Further, the review was limited to English-language publications, and requiring hypothesis-testing terms may also have limited retrieval of studies focused primarily on estimation. The final search was conducted on 13 October 2025, so the review does not cover subsequent publications. Title-and-abstract screening and full-text eligibility assessment were conducted by Reviewer~1, with Reviewer~3 reviewing flagged records or full texts and having access to the initial assessment and rationale. Because this additional review focused on flagged records, errors in other eligibility decisions could have remained undetected. Lastly, the review uses evidence mapping and descriptive appraisal of reporting, and no formal risk-of-bias assessment was conducted. The synthesis also depends on consistent extraction and classification, particularly when distinguishing reported confidence intervals from evaluations of coverage and stated power targets from numerical evaluations of power.

\section{Data and code availability}

The extracted study data are provided in the Study-Level Extraction supplement. The papers contributing to the summary counts are listed in the Methods and Count-to-Paper Tables supplement. Code for the topic-modeling analysis is available from the
authors on request.

\section{AI assistance statement}

Large-language-model tools were used to support source checks, grammar checking, sentence rewriting, code development and checking, topic-label generation, and figure preparation. Reviewer~1 conducted the initial title-and-abstract screening without AI assistance.

During manuscript revision and validation, GPT-6 Astra (OpenAI; ultra reasoning setting) and Claude Opus 5.5 (Anthropic; high and extra-high effort settings) assisted Reviewer~1 in checking the reporting tables (Study-Level Extraction Tables~\ref*{tab:appq1}--\ref*{tab:appq4b2}) and manuscript against the available articles, and in checking consistency between tables and narrative claims. Reviewer~1 used these checks to investigate potential discrepancies and cross-reference Reviewer~2's comments. Tool-generated suggestions were treated as provisional and reviewed by Reviewer~1; unresolved extraction disagreements were referred to Reviewer~3. Final eligibility, extraction, and synthesis decisions remained with the human reviewers. The authors take full responsibility for verifying AI-assisted text, code, and source checks and for all final decisions.

\clearpage
\bibliographystyle{oup-abbrvnat}
\bibliography{reference}



\end{document}